\documentclass[a4paper,11pt]{article}
\usepackage{pos}
\usepackage{cleveref}

\newcommand{\tauon}{$\tau$-lepton}
\newcommand{\tauons}{$\tau$-leptons}
\newcommand{\nupyprop}{\texttt{nuPyProp}}
\newcommand{\nuSpaceSim}{\texttt{$\nu$SpaceSim}}

\title{Neutrino propagation through Earth: modeling uncertainties using {\texttt{nuPyProp}}}
 \ShortTitle{{\texttt{Monte Carlo simulations: nuPyProp}}}

\manuallySeparateAuthors
\author*[a]{Diksha Garg}
\author[a]{and Mary Hall Reno}
\author{on behalf of nuSpaceSim collaboration}

\affiliation[a]{University of Iowa,\\
  Iowa City, Iowa, USA}

\emailAdd{diksha-garg@uiowa.edu}
\emailAdd{mary-hall-reno@uiowa.edu}

\abstract{Using the Earth as a neutrino converter, tau neutrino fluxes from astrophysical point sources can be detected by tau-lepton-induced extensive air showers (EASs). 
Both muon neutrino and tau neutrino induced upward-going EAS signals can be detected by terrestrial, sub-orbital and satellite-based instruments. 
The sensitivity of these neutrino telescopes can be evaluated with the {\texttt{nuSpaceSim}} package, which includes the {\texttt{nuPyProp}} simulation package. 
The {\texttt{nuPyProp}} package propagates neutrinos ($\nu_\mu$, $\nu_\tau$) through the Earth to produce the corresponding charged leptons (muons and tau-leptons). 
We use {\texttt{nuPyProp}} to quantify the uncertainties from Earth density models, tau depolarization effects and photo-nuclear electromagnetic energy loss models in the charged lepton exit probabilities and their spectra. 
The largest uncertainties come from electromagnetic energy loss modeling, with as much as a 20-50\% difference between the models. 
We compare {\texttt{nuPyProp}} results with other simulation package results.

}

\ConferenceLogo{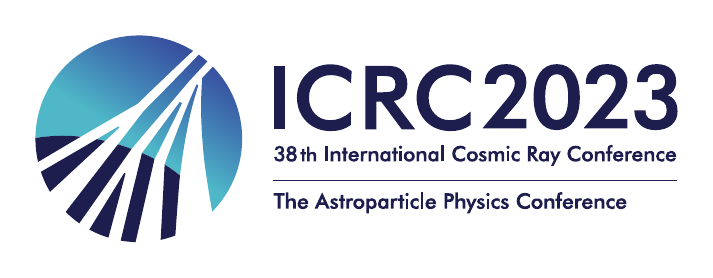}

\FullConference{%
The 38th International Cosmic Ray Conference (ICRC2023)\\
  26 July -- 3 August, 2023\\
  Nagoya, Japan}

\begin{document}
\maketitle

\section{Introduction}
Over the last several decades, astronomers and physicists have collaborated together to study the messengers of the universe: photons, neutrinos, cosmic rays, and gravitational waves. Ultra-high-energy cosmic rays (UHECRs) ($E > 10^{18}$ eV) are highly energetic particles composed of protons and nuclei, constantly pelting Earth.~They interact with Earth's atmosphere to produce a shower of other energetic particles.~Because they are charged particles, they are deflected by the magnetic fields that exists in the Universe while they travel to the Earth. 
UHECRs can interact with matter within astrophysical sources to produce charged pions which decay to produce very-high-energy (VHE) ($E>10^{15}$ eV) neutrinos. Neutrinos are also produced as UHECRs transit the Universe and interact with the cosmic photon background, again producing charged pions that decay.
Neutrinos, being neutral and weakly interacting, won't interact with matter or get deflected by magnetic fields on their way to the Earth. Thus, by studying these UHE neutrinos we can better understand the sources, evolution and composition of UHECRs, and also find the sources of the most energetic environment in the Universe. 

Beginning with cosmic ray production of charged pions, a series of decays, e.g., $\pi^+ \to \mu^+ + \nu_\mu$; $\mu^+ \to \bar\nu_\mu + \nu_e + e^+$ yields neutrinos.
The initial ratio of neutrino flavours at sources is disproportionate, ${N}_{\nu_e}:{N}_{\nu_\mu}:{N}_{\nu_\tau} \sim 1:2:0$. Production of $\nu_\tau$ flavour is highly suppressed at the source, but due to flavour mixings, neutrino oscillations over astronomical distances yields neutrino flavours arrive at Earth~\cite{Bustamante:2019sdb,Song:2020nfh} in proportion ${N}_{\nu_e}:{N}_{\nu_\mu}:{N}_{\nu_\tau} \sim 1:1:1$. 

\begin{figure*}[t]
    \centering
    \includegraphics[width=0.7\textwidth]{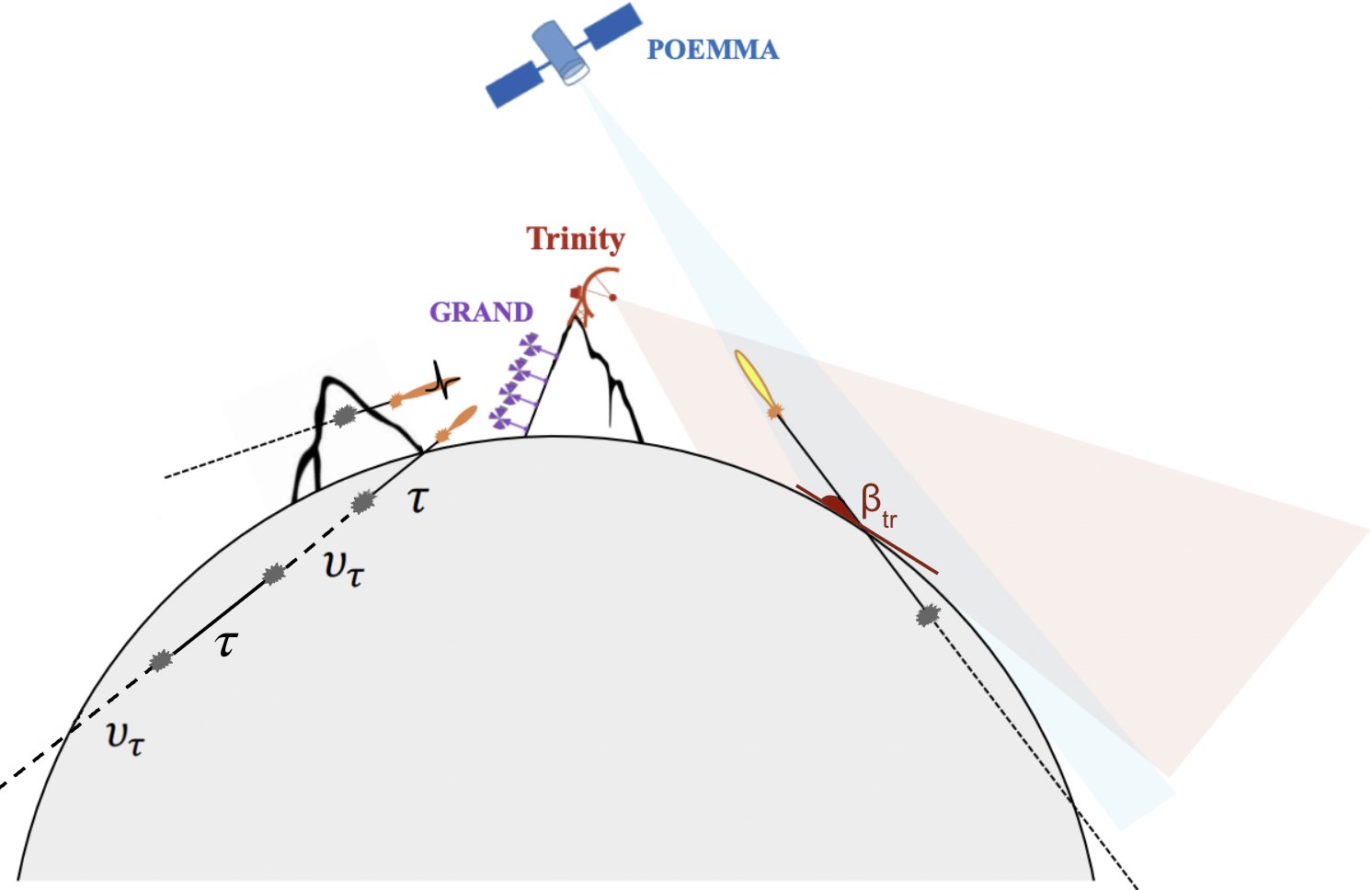}
    \caption{$\nu_\tau$ propagation through the Earth, producing {\tauon} which is exiting the Earth at $\beta_{tr}$ Earth emergence angle. EASs detected by neutrino experiments GRAND, TRINITY, and POEMMA. Figure reproduced from ref.~\cite{Huang:2021mki}.}
    \label{fig:neutrino_detec}
\end{figure*}

One of the methods of detecting UHE neutrinos is to use the Earth as a neutrino converter. Neutrinos of energy more than 40 TeV have high probability of interacting while propagating through the Earth.~One of the channels that is used by current and future neutrino experiments (GRAND~\cite{Alvarez-Muniz:2018bhp}, Trinity~\cite{Otte:2019aaf}, POEMMA~\cite{Olinto_2021}, etc.)~is $\nu_\tau$ propagation through the Earth, interacting to produce a {\tauon} which can exit the Earth at an emergence angle $\beta_{tr}$, and decay in the atmosphere to create an upward-going Extensive Air Shower (EAS)~\cite{PhysRevD.102.123013, Reno:2019jtr}. This channel is of interest because electromagnetic energy loss of {\tauons} in transit through the Earth is smaller than muons, because of both the higher mass of the \tauon\ and its shorter lifetime. When the {\tauons} decay, they regenerate tau-neutrinos which can again interact via the charged-current (CC) process to produce a {\tauon}. This is shown in~\cref{fig:neutrino_detec}. High energy regeneration processes are not a feature of $\nu_\mu$ propagation because muons have a long lifetime and many more electromagnetic loss interactions. When the muons finally decay, the decay $\nu_\mu$'s have low probability to interact with matter. 

To determine neutrino flux sensitivities of these experiments, there is a need for an end-to-end package to simulate the propagation of cosmic $\nu_\mu$ and $\nu_\tau$ through the Earth to produce EAS in the atmosphere. One such package is {\nuSpaceSim}~\cite{Krizmanic:2023icrc,Krizmanic:2021eyu}, designed to simulate radio and optical signals in the atmosphere that originate from $\nu_\tau$'s. The sensitivities of the experiments depends on the flux of {\tauons} and muons exiting the Earth, determined by the {\nupyprop}~\cite{Garg:2022ugd} simulation package, a standalone package that is part of {\nuSpaceSim} package. Using {\nupyprop}, the propagation through the Earth of $\nu_\mu$ and $\nu_\tau$, and the muons and \tauons\ they produce, yields charged lepton exit probabilities and energy distributions that do not depend on experiments, so {\nupyprop} is a mission independent simulation code. 

The next section gives an overview of the framework of {\nupyprop} and discusses the models/parametrizations used for neutrino/anti-neutrino cross-sections and electromagnetic energy loss cross-sections.~\Cref{sec:model_uncert} shows some selected results from {\nupyprop}.
More details and results appear in ref.~\cite{Garg:2022ugd}. 

\section{Framework and structure of {\nupyprop}} \label{sec:framework}
{\nupyprop} is a highly modular and flexible code which has only few library dependencies. It is an open-source code available on GitHub\footnote{\url{https://github.com/NuSpaceSim/nupyprop}}. Its main purpose is to simulate neutrinos ($\nu_\tau$, $\nu_\mu$, and anti-neutrinos) propagating through the Earth, interacting to produce charged leptons ({\tauons} and muons). It is designed for the energy range of E$_\nu = 10^6-10^{12}$ GeV. 
{\nupyprop} is coded in two languages, FORTRAN 90 which handles particle propagation and Python which does data handling. 

\begin{figure*}[t]
    \centering
    \includegraphics[width=0.9\textwidth]{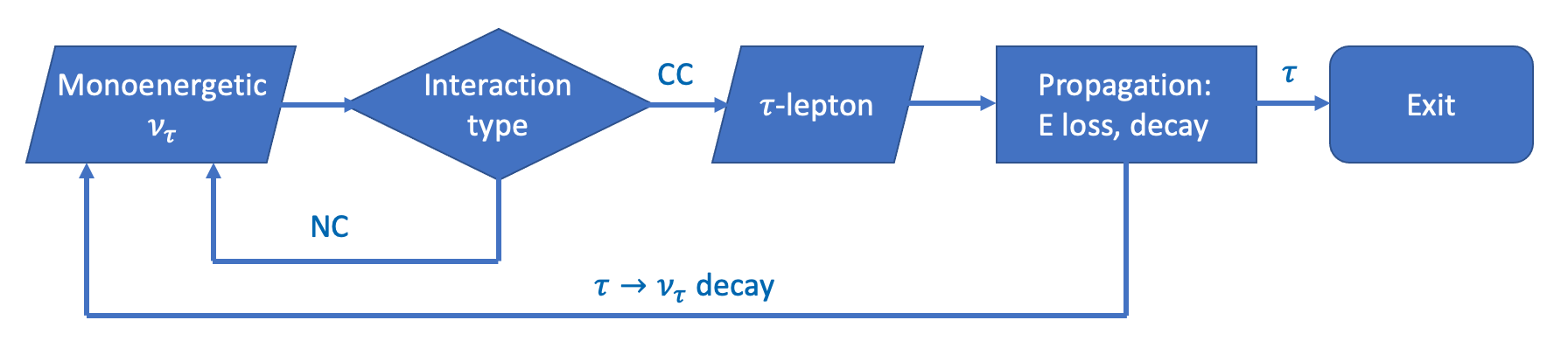}
    \caption{Flowchart explaining the particle propagation in {\nupyprop}. Figure reproduced from ref.~\cite{Garg:2022ugd}.}
    \label{fig:flowchart}
\end{figure*}

The working of {\nupyprop} is explained via the flowchart in~\cref{fig:flowchart}. We start with mono-energetic neutrinos which propagate through the Earth, first the water layer (the surface depth of the water layer can be set from $0-10$ km), then through rest of the Earth. If the neutrino interacts via a neutral current interaction, a lower energy neutrino is produced, and we are back at the beginning of the loop. If the neutrino interacts via a CC interaction, it produces the corresponding charged lepton. The charged lepton propagates through the Earth with electromagnetic interactions (ionization, bremsstrahlung, pair production, and photo-nuclear processes) that cause energy losses. If the charged lepton decays before exiting the Earth, it produces a lower energy (regenerated) neutrino, and we are back at the beginning of the loop. The regenerated neutrino plays an important role for {\tauons}. For charged leptons that exit the Earth,  {\nupyprop} generates output lookup tables which contains information on the exit probability, energy distributions, and average polarization of the exiting charged leptons. These output tables can be used as inputs for simulating EASs by packages like {\nuSpaceSim}.  

\begin{table}[]
\begin{tabular}{|c|c|}
    \hline
     \bf{Module} & \bf{Model/Parametrization/Type} \\ \hline
     Earth Density & PREM~\cite{DZIEWONSKI1981297} \\ \hline
     $\nu$/$\bar{\nu}$ Cross-Section & {\texttt{allm}~\cite{Abramowicz:1991xz, Dutta:2000hh}, \texttt{bdhm}~\cite{Block:2014kza}, \texttt{nct15}~\cite{Kovarik:2015cma}, \texttt{ct18nlo}~\cite{Hou:2019efy}, \texttt{ctw}~\cite{Connolly:2011vc} }  \\ \hline
     Lepton Photo-Nuclear  & \texttt{allm}~\cite{Abramowicz:1991xz,Abramowicz:1997ms}, \texttt{bdhm}~\cite{Block:2014kza}, \texttt{ckmt}~\cite{Capella:1994cr}, \texttt{bb}~\cite{Bezrukov:1981ci} \\ 
     interaction & \\ \hline
      Energy Loss Mechanism & Stochastic, Continuous      \\ \hline
\end{tabular}
\caption{\label{table:models} Input lookup table parameters used in {\nupyprop} for simulating particle propagation through Earth. }
\end{table}

The {\nupyprop} code requires millions of neutrinos propagating through the Earth to get good statistics for the output lookup tables. To reduce the computational time, {\nupyprop} does interpolations using the input lookup tables that contain Earth trajectories (based on Earth density), neutrino/anti-neutrino cross-sections and energy distributions, and electromagnetic energy loss interactions and energy distributions. The input lookup tables are made using the models shown in~\cref{table:models}. 

\section{Results and Discussion}\label{sec:model_uncert}

In this section, we compare the results obtained from varying different parameters and models used in {\nupyprop} to get information on the uncertainties arising from different input parameters. We vary: 1) the depth of the water layer around Earth, 2) the Earth density models, and 3) the electromagnetic interaction models. Additionally, we study the effects of {\tauon} depolarization. Furthermore, we also show a comparison between {\nupyprop} and other Earth propagating neutrino codes. The results shown here use default parameters of {\nupyprop}~\cite{Garg:2022ugd}, unless otherwise specified. 

\begin{figure*}[t]
    \centering
    \includegraphics[width=0.5\textwidth]{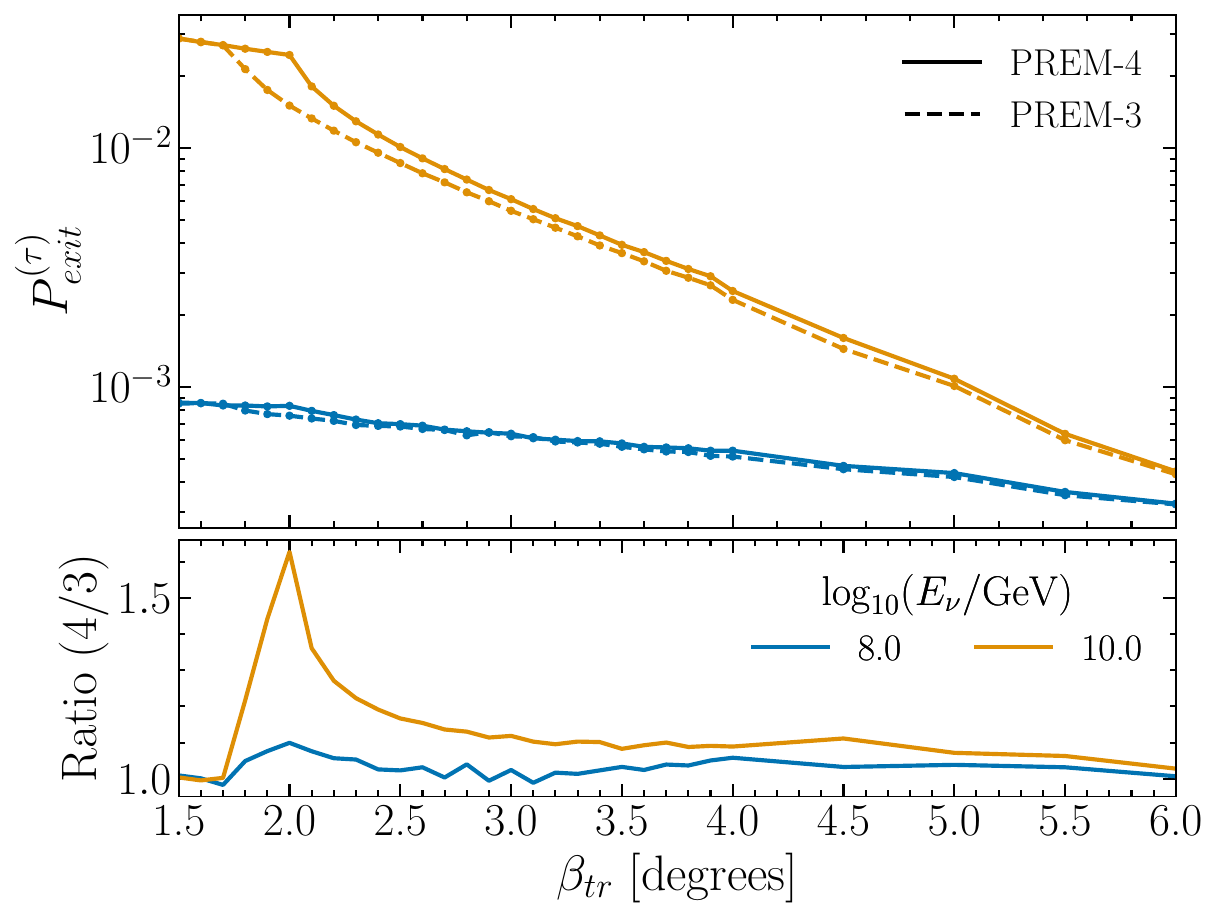}
    \includegraphics[width=0.47\textwidth]{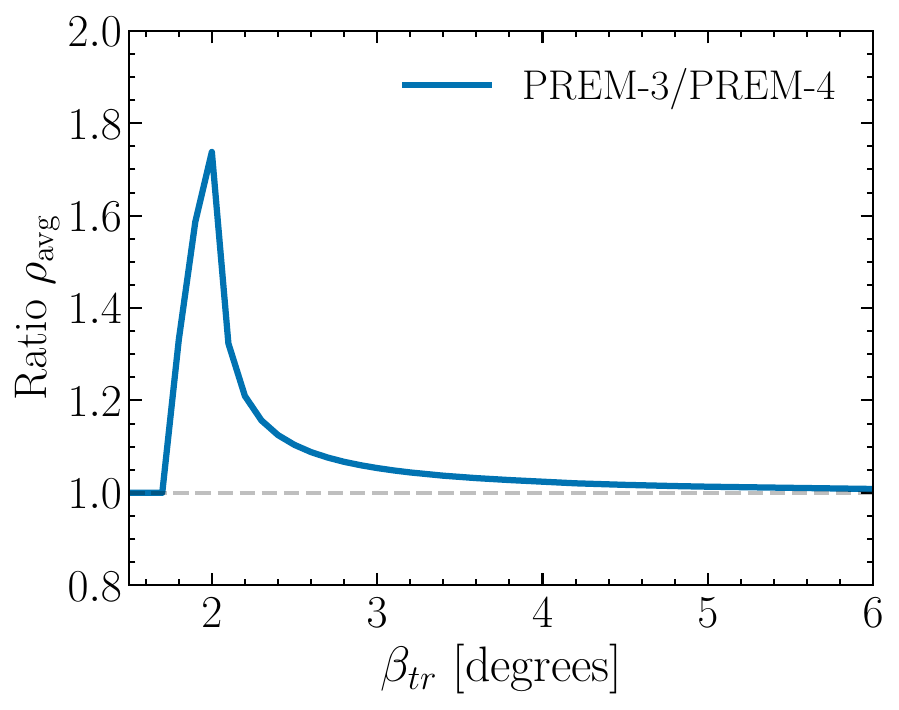}
    \caption{Left: Exit probability of {\tauon} as a function of $\beta_{tr}$ for PREM model for 3 and 4 km depth of water layer. Right: Ratio of average density of Earth for PREM model with 3 and 4 km water depth. Figure reproduced from ref.~\cite{Garg:2022ugd}. }
    \label{fig:pexit_prem}
\end{figure*}
\begin{figure*}[t]
    \centering
    \includegraphics[width=0.46\textwidth]{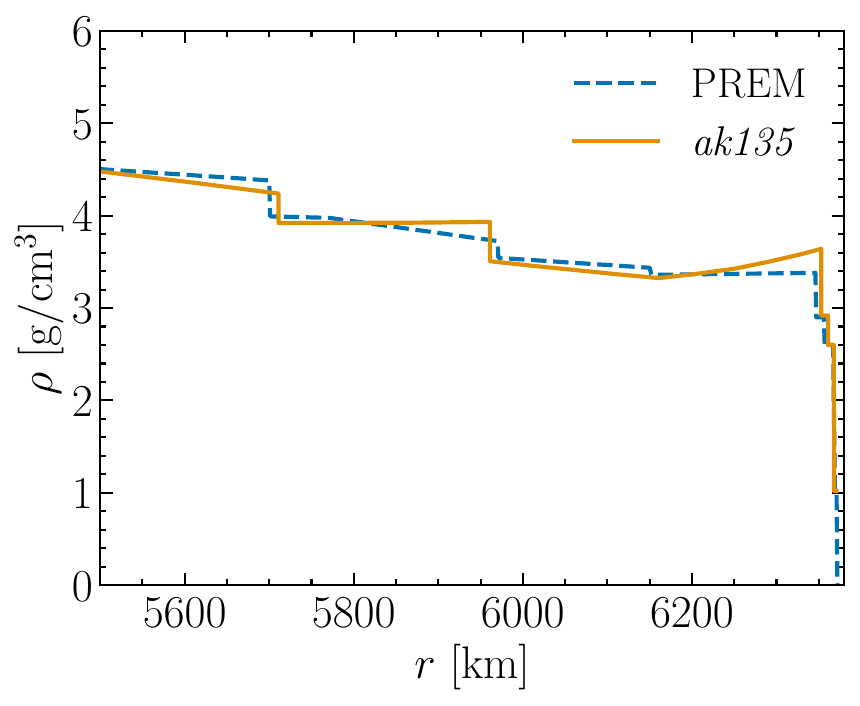}
    \includegraphics[width=0.48
    \textwidth]{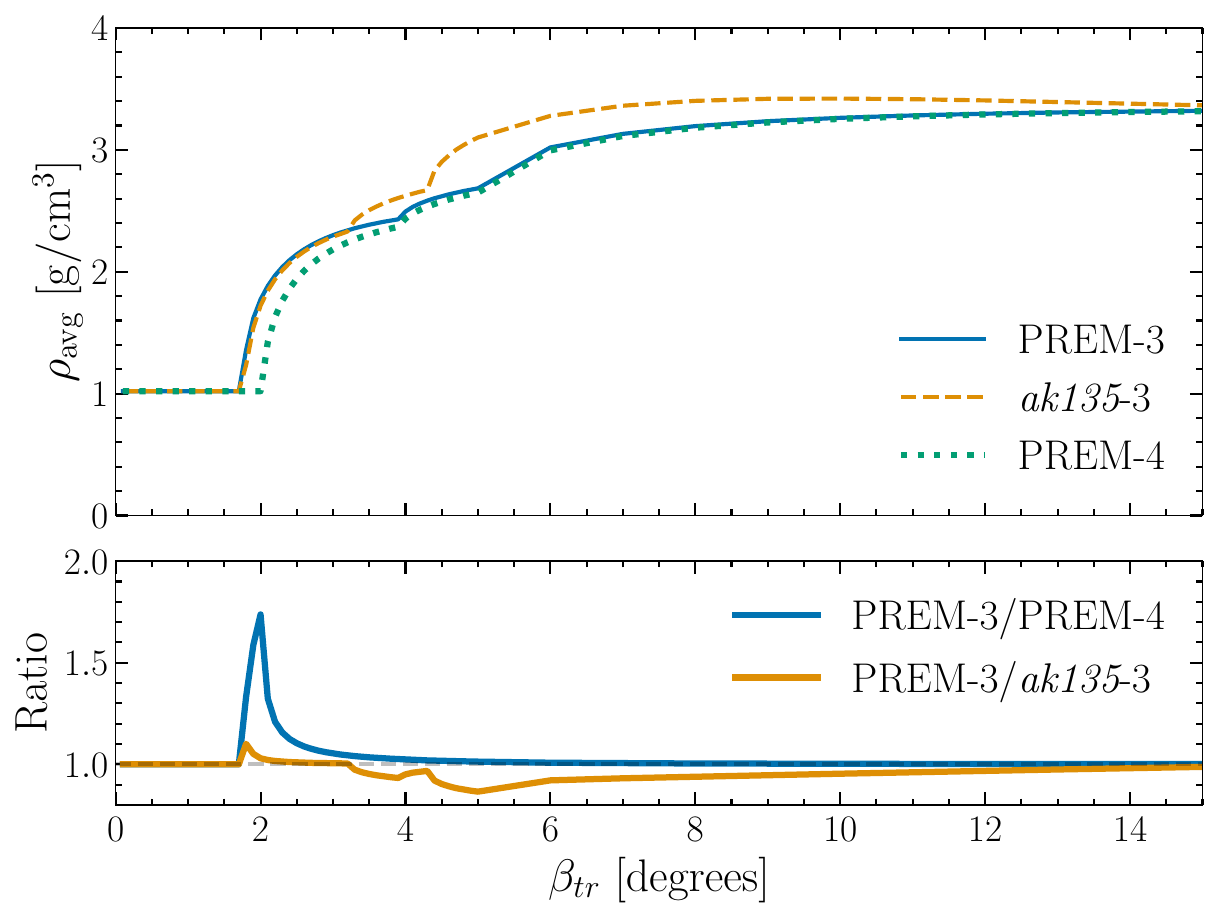} 
    \caption{Left: Earth density for PREM and {\it ak135} model as a function of radial distance from center of Earth ($r$). Right: Average density of Earth for PREM model with 3 and 4 km depth, and {\it ak135} model with 3 km depth of water layer. Figure reproduced from ref.~\cite{Garg:2022ugd}.}
    \label{fig:avgdensity}
\end{figure*}

We show the exit probability of {\tauons} as a function of Earth emergence angles in~\cref{fig:pexit_prem} (left), and we compare PREM Earth density models of water layer depths of 3 km  (PREM-3) and the {\nupyprop} default of 4 km (PREM-4) (right). The difference in \tauon\ exit probabilities between the two cases is mainly seen for $\beta_{tr}=1.7^{\circ}-2.3^\circ$. The water-rock interface for 3 km water depth layer is at $\beta_{tr}=1.76^\circ$, and for 4 km water depth layer is at $\beta_{tr}=2.03^\circ$.~The particles traversing the 3 km water depth layer plus rock undergo more energy losses as they hit the rock layer, than the particles traversing the 4 km water depth layer plus rock, thus the exit probability shows a sharper decrease for smaller $\beta_{tr}$ for the 3 km water depth case. For $E_\nu=10^{8}$ GeV, the difference between the two cases is $\sim 10\%$ at the water-rock interface, whereas for $E_\nu=10^{10}$ GeV, it is about $\sim 60\%$. The shape and approximate magnitude of the ratio of the exit probability is similar to the inverse of the ratio of average Earth density for PREM model with 3 km and 4 km water depth layer shown in~\cref{fig:pexit_prem} (right).

The Earth's density as a function of radial distance to the center of the Earth is updated in the {\it ak135}~\cite{Kennett:1995ak} Earth density model, which used an improved analysis of seismic wave data as compared to the PREM model. A comparison between the two models is shown in~\cref{fig:avgdensity} (left). Comparisons of the average Earth density as a function of $\beta_{tr}$ for the PREM models with 3 and 4 km water depth layers and the {\it ak135} model with 3 km water depth layer (labelled {\it ak135}-3) are shown in~\cref{fig:avgdensity} (right). The difference in average Earth density between PREM-3 and {\it ak135}-3 is of order $\sim 10-15\%$. Analogous to the impact of the water depth layer on the exit probability of {\tauons}, different Earth density models have $\sim 10-15\%$ differences on the exit probabilities of {\tauons}. 
One conclusion is that the main Earth density effect is at the water-rock interface and comes from the depth of the water layer considered. 

\begin{figure*}[t]
    \centering
    \includegraphics[width=0.6\textwidth]{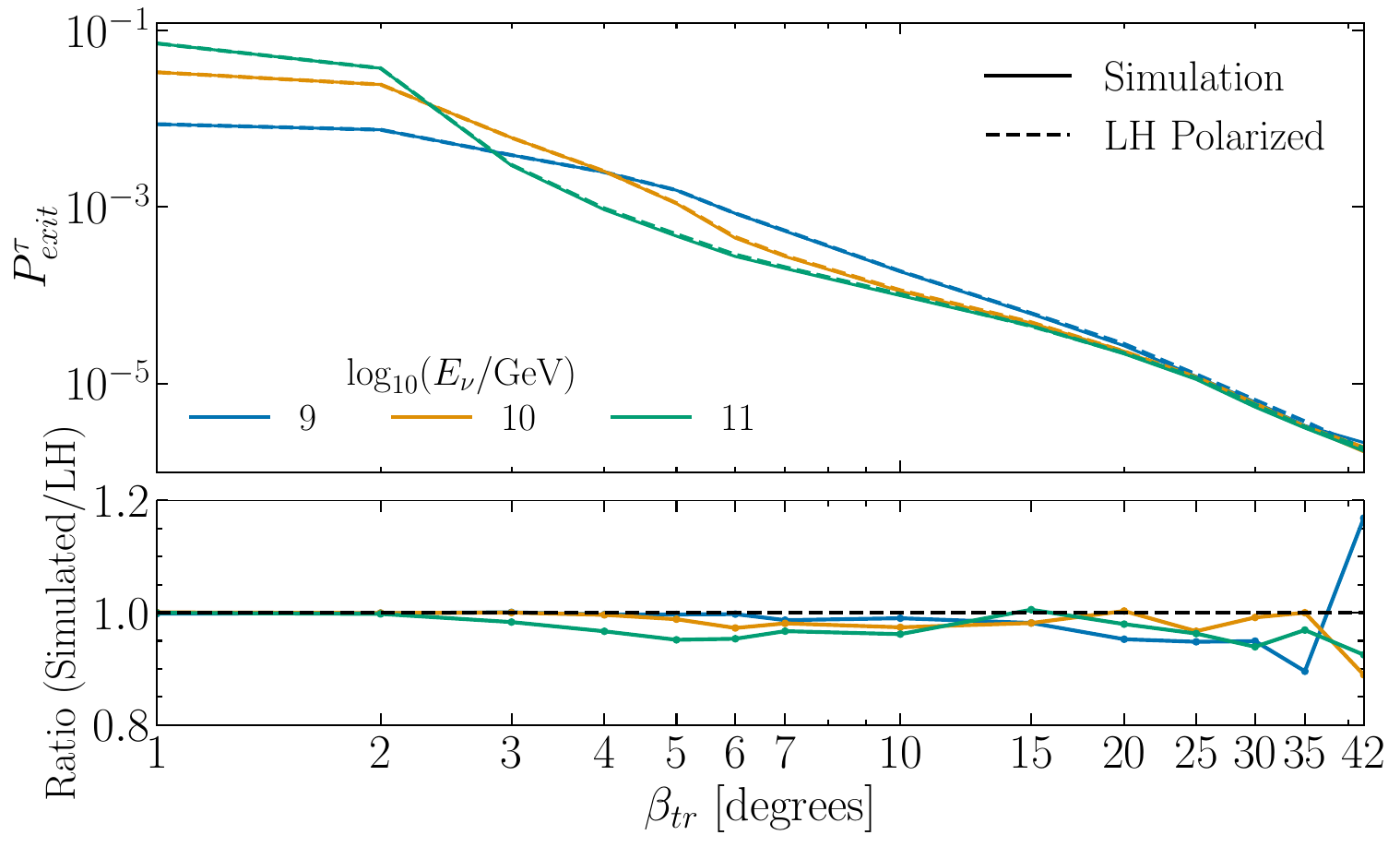}
    \caption{ Exit probability of {\tauon} as a function of $\beta_{tr}$ for simulated (solid) and LH (dashed) polarized case. Figure reproduced from ref.~\cite{Garg:2022ugd}.}
    \label{fig:pexit_polarization}
\end{figure*}

At VHE, the {\tauon} produced from the $\nu_\tau$ is 100\% polarized \cite{Arguelles:2022bma}, but with subsequent electromagnetic interactions, the {\tauon} can get depolarized. This is potentially important to study because the polarization value of {\tauons} impacts the regenerated neutrino energy distributions, as described in detail in refs.~\cite{Arguelles:2022bma,Garg:2022ugd}. The exit probabilities of {\tauons} are shown in~\cref{fig:pexit_polarization} for two cases: the simulated case where the depolarization of {\tauon} is considered for every photo-nuclear electromagnetic interaction; and for left-handed (LH) polarized case where the {\tauon} is considered to be LH for all electromagnetic interactions. Depolarization has a  $\sim 5\%$ effect for small Earth emergence angles, and a $\sim 10\%$ effect for larger angles. Overall, the depolarization effect on the exit probability of {\tauons} is small. 

\begin{figure*}[t]
    \centering
    \includegraphics[width=0.49\textwidth]{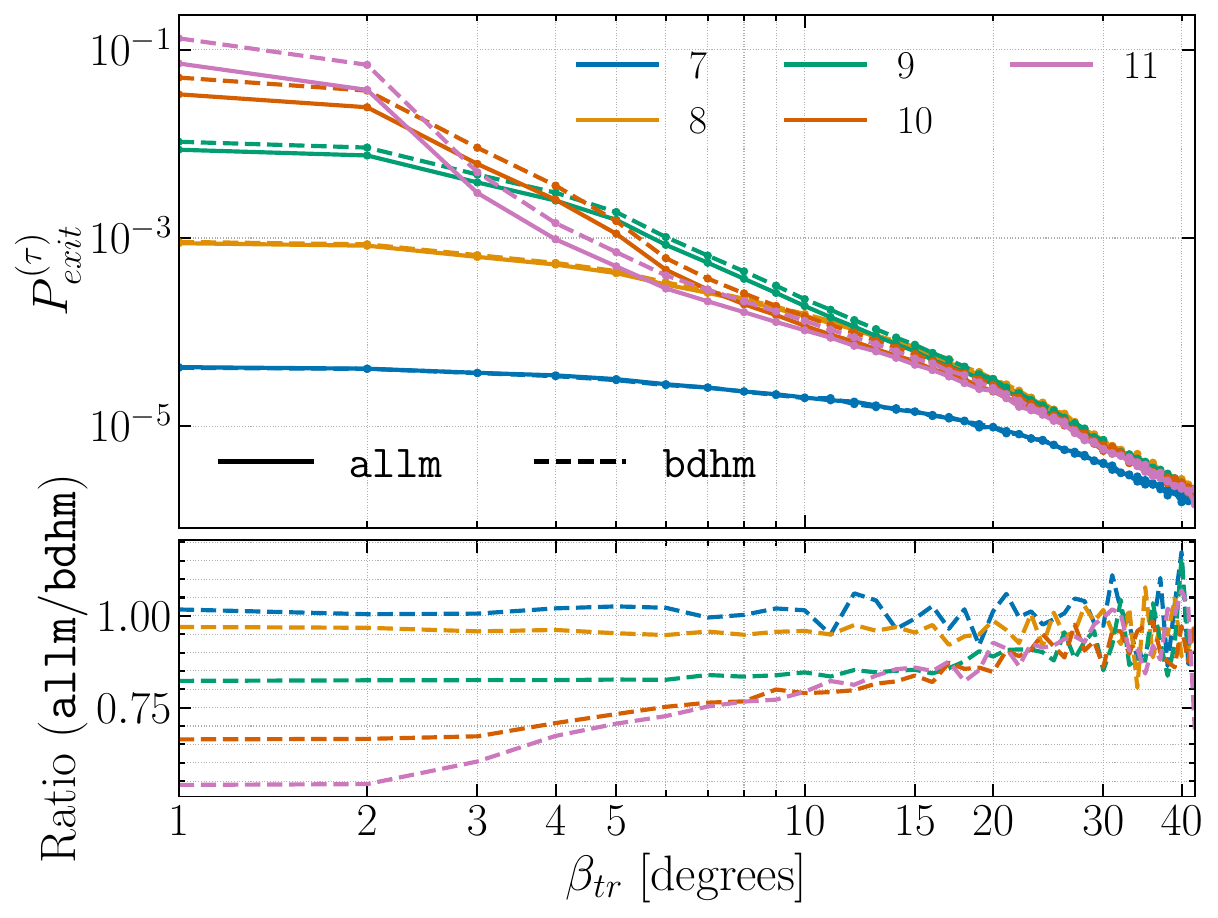}
    \includegraphics[width=0.49\textwidth]{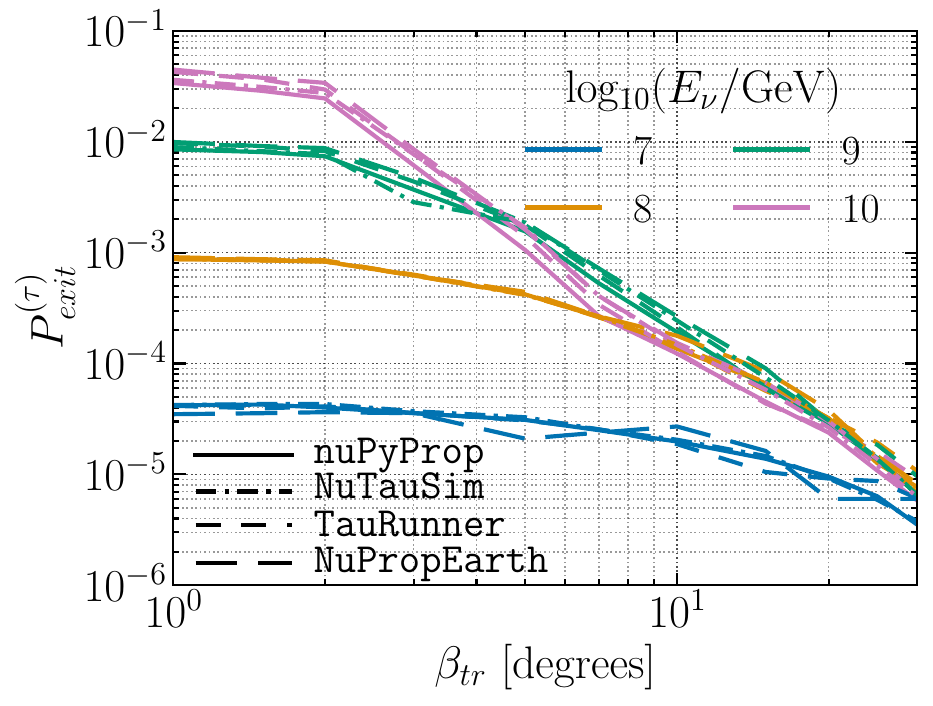}
    \caption{Left: Exit probability of {\tauon} as a function of $\beta_{tr}$ for \texttt{allm} and \texttt{bdhm} models for electromagnetic interactions. Right: Exit probability of {\tauon} as a function of $\beta_{tr}$ for different Monte-Carlo simulation codes. Figure reproduced from ref.~\cite{Garg:2022ugd}.}
    \label{fig:pexit_allm}
\end{figure*}

The {\tauon} electromagnetic energy loss modeling has a larger impact on \tauon\ exit probabilities than any other modeling done in {\nupyprop}.
The exit probabilities of {\tauons} as a function of $\beta_{tr}$ for two different electromagnetic energy loss models, \texttt{allm} and \texttt{bdhm}, are shown in~\cref{fig:pexit_allm} (left). The lower panel shows the ratio of the exit probabilities for the two models for different initial tau-neutrino energies. For $E_\nu=10^7$ GeV, the ratio is about unity but it starts decreasing for higher energies. This is because the energy loss parameter for photo-nuclear interaction for \texttt{allm} model is bigger than for \texttt{bdhm} model as shown in ref.~\cite{Garg:2022ugd}, and so the {\tauons} exiting the Earth are fewer under the \texttt{allm} evaluation than for \texttt{bdhm} evaluation. 
This accounts for the largest uncertainty in the \tauon\ exit probabilities of about $\sim 20-50\%$ for $E_\nu \ge 10^{9}$ GeV.~It arises from the extrapolations of the $F_2$ electromagnetic structure function for small-$x$ and large $Q^2$ region ($x$ is the fraction of nucleon's momentum carried by the struck quark and $Q^2$ is the momentum carried by the virtual photon in an electromagnetic interaction). 

Other Monte-Carlo simulation codes that propagate neutrinos through the Earth include \texttt{NuTauSim}~\cite{Alvarez-Muniz:2017mpk}, \texttt{TauRunner}~\cite{Safa:2021ghs}, and \texttt{NuPropEarth}~\cite{Garcia:2020jwr}. A detailed comparison of these codes is in ref.~\cite{Abraham:2022jse}. A comparison of these codes with {\nupyprop} on the exit probability of {\tauons} is shown in~\cref{fig:pexit_allm} (right). There is a good agreement between different codes across various tau-neutrino energies and angles. 

As we have shown here, the flexibility and modularity of {\nupyprop} allows users to import different input parameters which makes it easy to make comparisons between them and understand the uncertainties arising from different models. 

\section*{Acknowledgements}
This work is supported by NASA grants 80NSSC19K0626 at the University of Maryland, Baltimore County, 80NSSC19K0460 at the Colorado School of Mines, 80NSSC19K0484 at the University of Iowa, and 80NSSC19K0485 at the University of Utah, 80NSSC18K0464 at Lehman College, and RTOP 17-APRA17-0066 at NASA/GSFC and JPL.

\bibliographystyle{JHEP-nt}
\bibliography{references}

\providecommand{\href}[2]{#2}\begingroup\raggedright\begin{thebibliography}{10}

\bibitem{Bustamante:2019sdb}
M.~Bustamante and M.~Ahlers
  ~\href{https://doi.org/10.1103/PhysRevLett.122.241101}{\emph{Phys. Rev.
  Lett.} {\bfseries 122} (2019) 241101}
  [\href{https://arxiv.org/abs/1901.10087}{{\ttfamily 1901.10087}}].

\bibitem{Song:2020nfh}
N.~Song, S.W.~Li et~al.
  ~\href{https://doi.org/10.1088/1475-7516/2021/04/054}{\emph{JCAP} {\bfseries
  04} (2021) 054} [\href{https://arxiv.org/abs/2012.12893}{{\ttfamily
  2012.12893}}].

\bibitem{Alvarez-Muniz:2018bhp}
{\scshape GRAND} collaboration
  ~\href{https://doi.org/10.1007/s11433-018-9385-7}{\emph{Sci. China Phys.
  Mech. Astron.} {\bfseries 63} (2020) 219501}
  [\href{https://arxiv.org/abs/1810.09994}{{\ttfamily 1810.09994}}].

\bibitem{Otte:2019aaf}
A.N.~Otte, A.M.~Brown et~al. ~
  \href{https://arxiv.org/abs/1907.08727}{{\ttfamily 1907.08727}}.

\bibitem{Olinto_2021}
A.~Olinto et~al.
  ~\href{https://doi.org/10.1088/1475-7516/2021/06/007}{\emph{Journal of
  Cosmology and Astroparticle Physics} {\bfseries 2021} (2021) 007}.

\bibitem{PhysRevD.102.123013}
T.M.~Venters, M.H.~Reno, J.F.~Krizmanic et~al.
  ~\href{https://doi.org/10.1103/PhysRevD.102.123013}{\emph{Phys. Rev. D}
  {\bfseries 102} (2020) 123013}.

\bibitem{Reno:2019jtr}
M.H.~Reno, J.F.~Krizmanic et~al.
  ~\href{https://doi.org/10.1103/PhysRevD.100.063010}{\emph{Phys. Rev. D}
  {\bfseries 100} (2019) 063010}
  [\href{https://arxiv.org/abs/1902.11287}{{\ttfamily 1902.11287}}].

\bibitem{Krizmanic:2023icrc}
J.F.~{Krizmanic} et~al. ~\href{https://doi.org/10.22323/1.444.1110}{\emph{PoS}
  {\bfseries ICRC2023} (2023) 1110}.

\bibitem{Krizmanic:2021eyu}
{\scshape nuSpaceSim} collaboration
  ~\href{https://doi.org/10.22323/1.395.1205}{\emph{PoS} {\bfseries ICRC2021}
  (2021) 1205}.

\bibitem{Garg:2022ugd}
D.~Garg et~al.
  ~\href{https://doi.org/10.1088/1475-7516/2023/01/041}{\emph{JCAP} {\bfseries
  01} (2023) 041} [\href{https://arxiv.org/abs/2209.15581}{{\ttfamily
  2209.15581}}].

\bibitem{Huang:2021mki}
G.-y.~Huang, S.~Jana, M.~Lindner and W.~Rodejohann ~
  \href{https://arxiv.org/abs/2112.09476}{{\ttfamily 2112.09476}}.

\bibitem{DZIEWONSKI1981297}
A.M.~Dziewonski and D.L.~Anderson ~{\emph{Physics of the Earth and Planetary
  Interiors} {\bfseries 25} (1981) 297 }.

\bibitem{Abramowicz:1991xz}
H.~Abramowicz, E.M.~Levin, A.~Levy and U.~Maor
  ~\href{https://doi.org/10.1016/0370-2693(91)90202-2}{\emph{Phys. Lett.}
  {\bfseries B269} (1991) 465}.

\bibitem{Dutta:2000hh}
S.I.~Dutta, M.H.~Reno, I.~Sarcevic and D.~Seckel
  ~\href{https://doi.org/10.1103/PhysRevD.63.094020}{\emph{Phys. Rev.}
  {\bfseries D63} (2001) 094020}
  [\href{https://arxiv.org/abs/hep-ph/0012350}{{\ttfamily hep-ph/0012350}}].

\bibitem{Block:2014kza}
M.M.~Block, L.~Durand and P.~Ha
  ~\href{https://doi.org/10.1103/PhysRevD.89.094027}{\emph{Phys. Rev.}
  {\bfseries D89} (2014) 094027}
  [\href{https://arxiv.org/abs/1404.4530}{{\ttfamily 1404.4530}}].

\bibitem{Kovarik:2015cma}
K.~Kovarik et~al.
  ~\href{https://doi.org/10.1103/PhysRevD.93.085037}{\emph{Phys. Rev. D}
  {\bfseries 93} (2016) 085037}
  [\href{https://arxiv.org/abs/1509.00792}{{\ttfamily 1509.00792}}].

\bibitem{Hou:2019efy}
T.-J.~Hou et~al.
  ~\href{https://doi.org/10.1103/PhysRevD.103.014013}{\emph{Phys. Rev. D}
  {\bfseries 103} (2021) 014013}
  [\href{https://arxiv.org/abs/1912.10053}{{\ttfamily 1912.10053}}].

\bibitem{Connolly:2011vc}
A.~Connolly, R.S.~Thorne and D.~Waters
  ~\href{https://doi.org/10.1103/PhysRevD.83.113009}{\emph{Phys. Rev. D}
  {\bfseries 83} (2011) 113009}
  [\href{https://arxiv.org/abs/1102.0691}{{\ttfamily 1102.0691}}].

\bibitem{Abramowicz:1997ms}
H.~Abramowicz and A.~Levy ~
  \href{https://arxiv.org/abs/hep-ph/9712415}{{\ttfamily hep-ph/9712415}}.

\bibitem{Capella:1994cr}
A.~Capella, A.~Kaidalov, C.~Merino and J.~Tran Thanh~Van
  ~\href{https://doi.org/10.1016/0370-2693(94)90988-1}{\emph{Phys. Lett.}
  {\bfseries B337} (1994) 358}
  [\href{https://arxiv.org/abs/hep-ph/9405338}{{\ttfamily hep-ph/9405338}}].

\bibitem{Bezrukov:1981ci}
L.B.~Bezrukov and E.V.~Bugaev ~{\emph{Yad. Fiz.} {\bfseries 33} (1981) 1195}.

\bibitem{Kennett:1995ak}
B.L.N.~Kennett, E.R.~Engdahl and R.~Buland ~{\emph{Geophys. J. Int.} {\bfseries
  122} (1995) 108}.

\bibitem{Arguelles:2022bma}
C.A.~Arg\"uelles, D.~Garg et~al.
  ~\href{https://doi.org/10.1103/PhysRevD.106.043008}{\emph{Phys. Rev. D}
  {\bfseries 106} (2022) 043008}
  [\href{https://arxiv.org/abs/2205.05629}{{\ttfamily 2205.05629}}].

\bibitem{Alvarez-Muniz:2017mpk}
J.~Alvarez-Muniz et~al.
  ~\href{https://doi.org/10.1103/PhysRevD.97.023021}{\emph{Phys. Rev.}
  {\bfseries D97} (2018) 023021}
  [\href{https://arxiv.org/abs/1707.00334}{{\ttfamily 1707.00334}}].

\bibitem{Safa:2021ghs}
I.~Safa, J.~Lazar et~al. ~ \href{https://arxiv.org/abs/2110.14662}{{\ttfamily
  2110.14662}}.

\bibitem{Garcia:2020jwr}
A.~Garcia, R.~Gauld, A.~Heijboer and J.~Rojo
  ~\href{https://doi.org/10.1088/1475-7516/2020/09/025}{\emph{JCAP} {\bfseries
  09} (2020) 025} [\href{https://arxiv.org/abs/2004.04756}{{\ttfamily
  2004.04756}}].

\bibitem{Abraham:2022jse}
R.M.~Abraham et~al. ~ \href{https://arxiv.org/abs/2203.05591}{{\ttfamily
  2203.05591}}.

\end{thebibliography}\endgroup

\clearpage
\section*{Full Authors List: nuSpaceSim collaboration}


\scriptsize
\noindent
Sameer Patel$^1$,
Alexander Ruestle$^2$,
Yosui Akaike$^3$,
Luis A. Anchordoqui$^4$,
Douglas R. Bergman$^5$,
Isaac Buckland$^5$,
Austin L. Cummings$^{6,7}$,
Johannes Eser$^8$,
Fred Angelo Batan Garcia$^{2,9}$,
Claire Gu\'epin$^{8,9}$,
Tobias Heibges$^6$,
Andrew Ludwig$^{10}$,
John F. Krizmanic$^2$,
Simon Mackovjak$^{11}$,
Eric Mayotte$^6$,
Sonja Mayotte$^6$,
Angela V. Olinto$^8$,
Thomas C. Paul$^4$,
Andr\'es Romero-Wolf$^{10}$,
Fr\'ed\'eric Sarazin$^6$,
Tonia M. Venters$^2$,
Lawrence Wiencke$^6$,
and
Stephanie Wissel$^7$ \\

\noindent

$^1${Department of Physics and Astronomy,
University of Iowa, Iowa City, IA 52242, USA},
$^2${Laboratory for Astoparticle Physics, NASA/Goddard Space Flight Center, Greenbelt, MD 20771, USA},
$^3${Waseda Research Institute for Science and Engineering, Waseda University, Tokyo 162-0044, Japan},
$^4${Department of Physics and Astronomy, Lehman College, City University of New York, Bronx, NY 10468, USA},
$^5${Department of Physics and Astronomy, University of Utah, Salt Lake City, UT 84112, USA},
$^6${Department of Physics, Colorado School of Mines, Golden, CO 80401, USA},
$^7${Department of Physics, Department of Astronomy and Astrophysics, Institute for Gravitation and the Cosmos, Pennsylvania State University, State College, PA 16801},
$^8${Department of Astronomy and Astrophysics, University of Chicago, Chicago, IL 60637, USA},
$^9${Department of Astronomy, University of Maryland, College Park, MD 20742, USA},
$^{10}${Jet Propulsion Laboratory, Pasadena, CA 91109, USA},
and
$^{11}${Department of Space Physics, Institute of Experimental Physics, Slovak Academy of Sciences, 040 01 Ko\v{s}ice, Slovakia}

\end{document}